\documentclass[twocolumn,showpacs,preprintnumbers,amsmath,amssymb,prb]{revtex4}

\usepackage{graphicx}
\usepackage{bm}

\usepackage{amsfonts}

\begin{document}
\title{Influence of the mean flow on zonal flow generation}
\author{Volodymyr M. Lashkin  }
\email{vlashkin@kinr.kiev.ua}
 \affiliation{Institute for Nuclear
Research, Pr. Nauki 47, Kiev 03680, Ukraine}

\date{\today}

\begin{abstract}
Excitation of zonal flow by the modulational instability in the
presence of mean shear flow is considered. It is shown that the
small amplitude mean flow favours the modulational instability,
increasing the growth rate, whereas sufficiently strong mean shear
significantly reduces the instability growth rate.
\end{abstract}

\maketitle

It is now widely recognized that zonal flows, i. e. azimuthally
symmetric band-like shear flows that depend only on the radial
coordinate \cite{Diamond05,DiamondRev2} play a crucial role in
regulating drift-wave turbulence and transport in tokamaks. It is
now quite clear that zonal flows are generated by modulational
instability of drift waves
\cite{Smol1,Smol2,Chen,Champ,Manfredi,Zonca}. It is important to
distinguish the zonal flow from the mean shear flow associated
with the mean radial electric field. The latter can be driven and
sustained in the absence of turbulence (by heating, fueling,
momentum input, etc.), whereas zonal flows are driven exclusively
by nonlinear wave interaction processes (via the modulational
instability) \cite{Diamond05,DiamondRev2}. In contrast to smooth
static mean flows, the zonal flow patterns have complex, possibly
random, spatial structure. It is well known that the presence of
mean flow give rise not only to instability of the sheared layer
(the Kelvin--Helmholtz instability), but also to stabilization of
other instabilities (ion temperature gradient driven modes,
resistive interchange modes etc.) \cite{Diamond05, Tajima,
Hamaguchi, Waltz, Sugama}. In this Brief Communication the
influence of the mean shear flow on excitation of zonal flow by
the modulational instability is reported.

We use a simple slab two-dimensional ($x$ and $y$ being the radial
and poloidal coordinates respectively) model similar to the
Hasegawa-Mima model, but taking into account that the plasma
density does not follow the Boltzmann distribution for large scale
motions \cite{Smol2,Chen,Champ,Manfredi}. We represent the
electrostatic potential as a sum of small scale fluctuating
$\tilde{\varphi}$ and mean $\langle\varphi\rangle$ quantities $
\varphi=\tilde{\varphi}+\langle\varphi\rangle $, where
$\langle\varphi\rangle=\hat{\varphi}+\varphi_{0}$ with
$\hat{\varphi}$ being the zonal flow potential and $\varphi_{0}$
corresponds to the mean shear flow. Equations describing the
coupled dynamics of large (zonal and mean flows) and small (drift
waves) scale motions
\begin{equation}
\label{main1} \frac{\partial}{\partial
t}(1-\Delta)\tilde{\varphi}+\frac{\partial\tilde{\varphi}}{\partial
y}-\langle\mathbf{v}\rangle\cdot\nabla
\Delta\tilde{\varphi}-\mathbf{\tilde{v}}\cdot\nabla\Delta(\hat{\varphi}+\varphi_{0})=0,
\end{equation}
\begin{equation}
\label{main2} \frac{\partial}{\partial
t}\Delta\hat{\varphi}+\langle\tilde{\mathbf{v}}\cdot\nabla\Delta\tilde{\varphi}\rangle=0,
\end{equation}
where $\mathbf{\tilde{v}}=\mathbf{z}\times\nabla\tilde{\varphi}$,
$\langle\mathbf{v}\rangle=v_{0}(x)\mathbf{y}+\mathbf{z}\times\nabla\hat{\varphi}
$ and $v_{0}(x)$ is a profile of the mean flow,
$\langle\dots\rangle$ denotes the averaging process. The
normalizations of space in unit of $\rho_{s}$, time in units of
$L_{n}/c_{s}$, and potentials in units of $T_{e}\rho_{s}/L_{n}e$
are used in Eqs. (\ref{main1}) and (\ref{main2}). Here notations
are standard, i.e. $T_{e}$ is the electron temperature,
$c_{s}=\sqrt{T_{e}/m_{i}}$, $\rho_{s}=c_{s}/\Omega_{i}$ is the
ion-sound Larmor radius,  and $L_{n}$ is the characteristic length
scale of the plasma inhomogeneity. In the linear approximation Eq.
(\ref{main1}) yields the ordinary dispersion relation for drift
waves
\begin{equation}
\omega_{\mathbf{k}}=\frac{k_{y}}{1+k^{2}}.
\end{equation}
The term in the angular brackets of Eq. (\ref{main2}) corresponds
to the Reynolds stress induced by the small scale drift waves.

The electrostatic potential $\tilde{\varphi}$ of the drift waves
is considered as a superposition of a pump wave and two
satellites,
$\tilde{\varphi}=\tilde{\varphi}_{d}+\tilde{\varphi}_{+}+\tilde{\varphi}_{-}$.
In what follows, we consider the case when the drift wave is
propagating nearly in the poloidal direction, $k_{x}=0$ and
represent
\begin{gather}
\hat{\varphi}=\varphi(x)e^{-i\Omega t}+\varphi^{\ast}(x)e^{i\Omega
t}, \label{r1} \\
\tilde{\varphi}_{d}=\Phi_{0}e^{i(k_{y}y-\omega_{\mathbf{k}}t)}+
\Phi_{0}^{\ast}e^{-i(k_{y}y-\omega_{\mathbf{k}}t)},\\
\tilde{\varphi}_{\pm}=\varphi_{\pm}(x)e^{i(k_{y}y-\omega_{\pm}t)}+
\varphi_{\pm}^{\ast}(x)e^{-i(k_{y}y-\omega_{\pm}t)}, \label{r2}
\end{gather}
where $\omega_{\pm}=\omega_{\mathbf{k}}\pm\Omega$. Substituting
Eqs. (\ref{r1})--(\ref{r2}) into Eq. (\ref{main2}), we obtain

\begin{equation}
\Omega\varphi=k_{y}\left(\Phi_{0}
\frac{d\varphi_{-}^{\ast}}{dx}-\Phi_{0}^{\ast}
\frac{d\varphi_{+}}{dx}\right). \label{e1}
\end{equation}
For the amplitudes of the satellites we have from Eq.
(\ref{main1})
\begin{gather}
\left\{\left[\Omega+\omega_{\mathbf{k}}-k_{y}v_{0}(x)\right]
\left(1+k_{y}^{2}-\frac{d^{2}}{dx^{2}}\right)-k_{y}-k_{y}v_{0}''\right\}\varphi_{+}
\nonumber \\
=-k_{y}\Phi_{0}\left(k_{y}^{2}\frac{d}{dx}+\frac{d^{3}}{dx^{3}}\right)\varphi,
\label{e2}
\end{gather}

\begin{gather}
\left\{\left[\Omega-\omega_{\mathbf{k}}-k_{y}v_{0}(x)\right]
\left(1+k_{y}^{2}-\frac{d^{2}}{dx^{2}}\right)+k_{y}+k_{y}v_{0}''\right\}\varphi_{-}^{\ast}
\nonumber \\
=k_{y}\Phi_{0}^{\ast}\left(k_{y}^{2}\frac{d}{dx}+\frac{d^{3}}{dx^{3}}\right)\varphi
\label{e3},
\end{gather}
where $v_{0}''=d^{2}v_{0}(x)/dx^{2}$. In the absence of mean shear
flow ($v_{0}=0$), taking $\varphi,\varphi_{+}\sim \exp (iqx)$,we
recover the previous dispersion relation \cite{Shukla1,Shukla2}
\begin{equation}
\Omega^{2}=\delta^{2}+\frac{2k_{y}^{2}(q^{2}-
k_{y}^{2})q^{2}|\Phi_{0}|^{2}}{1+k_{y}^{2}+q^{2}},
\end{equation}
where
\begin{equation}
\delta= \frac{k_{y}q^{2}}{(1+k_{y}^{2})(1+k_{y}^{2}+q^{2})},
\end{equation}
which predicts the modulational instability ($\Omega^{2}<0$) if
the drift wave amplitude $|\Phi_{0}|$ is above the corresponding
threshold. Under $q\ll k_{y}$ maximum growth rate is achieved at
$q_{opt}=k_{y}|\Phi_{0}|(1+k_{y}^{2})^{3/2}$ and $
\gamma_{max}=k_{y}^{3}|\Phi_{0}|^{2}(1+k_{y}^{2})$.

\begin{figure}
\includegraphics[width=3.4in]{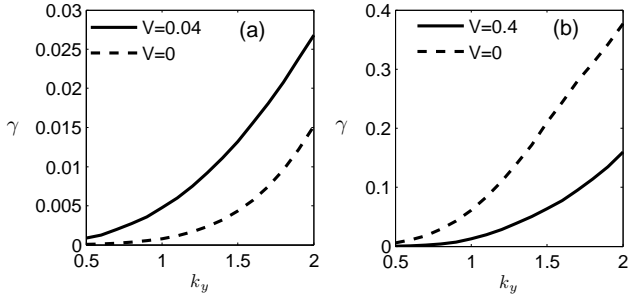} \caption{The
growth rate versus $k_{y}$ for $q_{0}=0.05$ and (a) mean flow
amplitude $V=0$ and $V=0.04$, drift wave amplitude
$|\Phi_{0}|=0.02$; (b) mean flow amplitude $V=0$ and $V=0.4$,
drift wave amplitude $|\Phi_{0}|=0.2$.} \label{fig1}
\end{figure}

\begin{figure}
\includegraphics[width=3.4in]{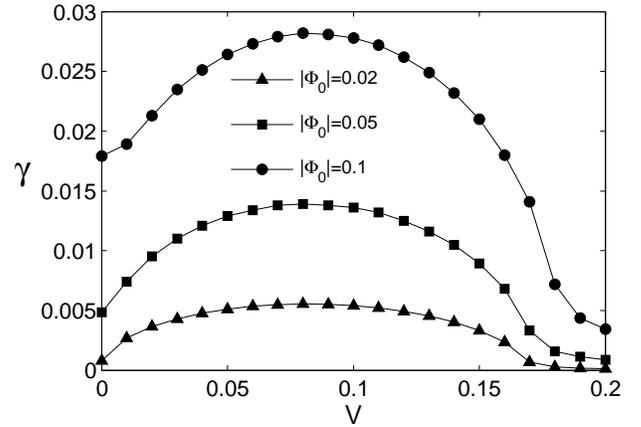} \caption{The
growth rate versus the mean flow amplitude $V$ for three different
values of the drift wave amplitude: $|\Phi_{0}|=0.02, 0.05, 0.1$.
Here $k_{y}=1$ and $q_{0}=0.05$.} \label{fig2}
\end{figure}

In the presence of mean flow ($v_{0}\neq 0$) equations
(\ref{e1})--(\ref{e3}) can be rewritten as the generalized
eigenvalue problem
\begin{equation}
\begin{pmatrix}
0 & A & -A \\
B & C & 0 \\
-B & 0 & D
\end{pmatrix}
\Psi
=\Omega \begin{pmatrix} 1 & 0 & 0 \\
0 & E & 0 \\
0 & 0 & E
\end{pmatrix}
\Psi \label{eigen}
\end{equation}
 where
$\Psi=(\varphi,\varphi_{+},\varphi_{-}^{\ast})$, and
\begin{gather}
A=-k_{y}\Phi_{0}\frac{d}{dx}, \quad
B=-k_{y}\Phi_{0}\left(k_{y}^{2}\frac{d}{dx}+\frac{d^{3}}{dx^{3}}\right),\\
C=k_{y}+k_{y}v_{0}''-(\omega_{\mathbf{k}}-k_{y}v_{0})E,\\
D=(\omega_{\mathbf{k}}+k_{y}v_{0})E-k_{y}-k_{y}v_{0}'',\\
E=1+k_{y}^{2}-\frac{d^{2}}{dx^{2}}.
\end{gather}
We assume that the profile of the mean shear flow has the form
$v_{0}(x)=V\tanh(q_{0} x)$. Employing a finite differencing
approximation, we numerically solved the eigenvalue problem
(\ref{eigen}). The instability corresponds to eigenvalues with
$\mathrm{Im}\,\Omega \neq 0$. Figures \ref{fig1}(a) and (b) show
the maximum growth rate $\gamma=\max |\mathrm{Im}\,\Omega |$ of
the modulational instability modified by the mean shear flow as a
function of the poloidal wave number $k_{y}$. Dashed lines
correspond to the case of the absence of mean flow. It is seen
that the small amplitude mean flow [Fig. \ref{fig1}(a)] favours
the modulational instability and zonal flow generation, increasing
the growth rate, whereas sufficiently strong mean shear [Fig.
\ref{fig1}(b)] significantly reduces the instability growth rate.
In Fig. \ref{fig2} we plot the growth rate for  fixed $k_{y}$ and
several values of the drift wave amplitude $|\Phi_{0}|$ as a
function of the mean flow amplitude $V$. The growth rate $\gamma$
initially increases with increasing $V$, and thus the presence of
mean flow has a destabilizing effect. Then, $\gamma$ attains its
maximum at some $V$ ($V\sim 0.08$ for $k_{y}=1$ and all
$|\Phi_{0}|$). This corresponds to the most unstable case. Then,
the growth rate decreases as $V$ increases, and when the mean flow
amplitude exceeds some critical value, the growth rate becomes
smaller then in the case of the absence of mean flow. Further
increasing of $V$ can significantly reduce the growth rate. Thus
the presence of mean flow with sufficiently strong amplitude has a
stabilizing effect on  generation of zonal flows by the
modulational instability.

 This work was  supported in part by Ukranian
Academy of Sciences through a programme 'Fundamental problems in
particle physics and nuclear energy', Grant No. 150/166.

\end{document}